**Title: Scaling laws for Light Absorption by Atmospheric Black Carbon Aerosol**

**Authors:** Rajan K. Chakrabarty[1,2]* and William R. Heinson[1]

**Affiliations:**

[1]Center for Aerosol Science and Engineering, Department of Energy, Environmental and Chemical Engineering, Washington University in St. Louis, Missouri – 63130, USA.

[2]McDonnell Center for the Space Sciences, Washington University in St. Louis, Missouri – 63130, USA.

* Correspondence to*: [chakrabarty@wustl.edu](mailto:chakrabarty@wustl.edu)*

**ABSTRACT**

Black carbon (BC) aerosol, the strongest absorber of visible solar radiation in the atmosphere, contributes to a large uncertainty in direct radiative forcing estimates. A primary reason for this uncertainty is inaccurate parameterizations of BC mass absorption cross-section ($MAC_{BC}$) and its enhancement factor ($E_{MAC_{BC}}$)–resulting from internal mixing with non-refractory and non-light absorbing materials–in climate models. Here, applying scaling theory to numerically-exact electromagnetic calculations of simulated BC particles and observational data on BC light absorption, we show that $MAC_{BC}$ and $E_{MAC_{BC}}$ evolve with increasing internal mixing ratios in simple power-law exponents of 1/3. Remarkably, $MAC_{BC}$ remains inversely proportional to wavelength at any mixing ratio. When mixing states are represented using mass-equivalent core-shell spheres, as is done in current climate models, it results in significant under prediction of $MAC_{BC}$. We elucidate the responsible mechanism based on shielding of photons by a sphere's skin depth and establish a correction factor that scales with a ¾ power-law exponent.

Black carbon (BC) aerosol, emitted from anthropogenic and natural combustion processes, dominates the absorption of incoming shortwave solar radiation in the earth's atmosphere and is considered the second largest contributor to global warming after carbon dioxide [1,2]. In spite of its perceived importance in climate change, there exists a large discrepancy between model- and observation-based estimates of direct radiative forcing (DRF) by BC. A primary reason for this discrepancy could be attributed to the systematic underestimation of light absorption by BC, up to a factor of 3, in climate models compared to observationally-constrained estimates [3,4]. Current efforts to address this disagreement have been directed toward scaling up of BC mass absorption cross-sections (MAC$_{BC}$)–defined as absorption cross-section per unit particle mass–in model parameterizations [1,5,6]. Accurate estimates of MAC$_{BC}$ are critical for enabling models convert measured or modeled BC mass concentrations over a region to representative absorption coefficients, which serve as input parameter to radiative transfer algorithms [1].

The upward scaling of MAC$_{BC}$ estimates in climate models has been motivated by recent field observations of BC existing in majority populations as internally mixed with non-refractory materials, including sulfate, nitrate, and organic carbon (OC) in the atmosphere [7-10]. Along with BC, combustion sources co-emit large amounts of volatile OC compounds, in addition to $NO_x$ and $SO_x$, which upon undergoing atmospheric processing condense on BC particle surfaces as layers of external coating. These layers, which are typically non-absorbing in the visible solar spectrum, acts as "focusing lens" for the incoming light and results in an enhanced absorption cross-section or MAC$_{BC}$ compared to that for an equivalent external mixture [9,11]. A broad range of enhancement factors for MAC$_{BC}$ (henceforth referred to as $E_{MAC_{BC}}$), from 1.05 to 3.5, has been observed during laboratory and field studies [8-10,12]. This large spread in $E_{MAC_{BC}}$ values accompanied by lack of any established scaling relationship makes it a cumbersome and challenging parameter to incorporate in models. A commonly adopted practice is, therefore, to either multiply uncoated $MAC_{BC}$ values with a constant $E_{MAC_{BC}}$ or estimate approximate $MAC_{BC}$ based on over-simplified aerosol models such as core-shell [6,13].

Here, we integrate results obtained from fractal modeling of internally-mixed BC and numerically-exact electromagnetic calculations with recent observational findings to establish universal scaling relationships for $E_{MAC_{BC}}$ and $MAC_{BC}$ as a function of BC coating mass in the shortwave solar wavelengths ($\lambda$ = 400 - 900 nm). Field and laboratory [7,8] data show that BC exists in the atmosphere primarily in three internally-mixed morphologies (figure 1): *bare aggregates* with point-contacting monomers, *partly coated aggregates* with monomer crevices filled with coating material but the aggregate not completely engulfed, and *embedded aggregates* with heavy coating mass and only the contours of the monomers evident. The dimensionless parameter $\frac{M_{total}}{M_{BC}}$ for a coated aggregate, defined as ratio of total particle mass (i.e., coating mass plus BC mass) to BC mass, ranges approximately between 1 and 5 for the partially coated, and $\geq$ 6 for the embedded types. Recent studies[8,12] have highlighted a particle's $\frac{M_{total}}{M_{BC}}$ as the primary parameter in controlling its $E_{MAC_{BC}}$. We show that when $E_{MAC_{BC}}$ and $MAC_{BC}$ datasets are analyzed systematically as a function of $\frac{M_{total}}{M_{BC}}$, remarkable power-law scaling relations of the form $Y = Y_0 S^\beta$, where Y is either $E_{MAC_{BC}}$ or $MAC_{BC}$, S is either $\frac{M_{total}}{M_{BC}}$ or $\lambda$, and $Y_0$ is the prefactor and $\beta$ is the power-law exponent, emerge. These simple scale dependencies would enable climate models to accurately and inexpensively incorporate the absorption properties of BC in their parameterizations.

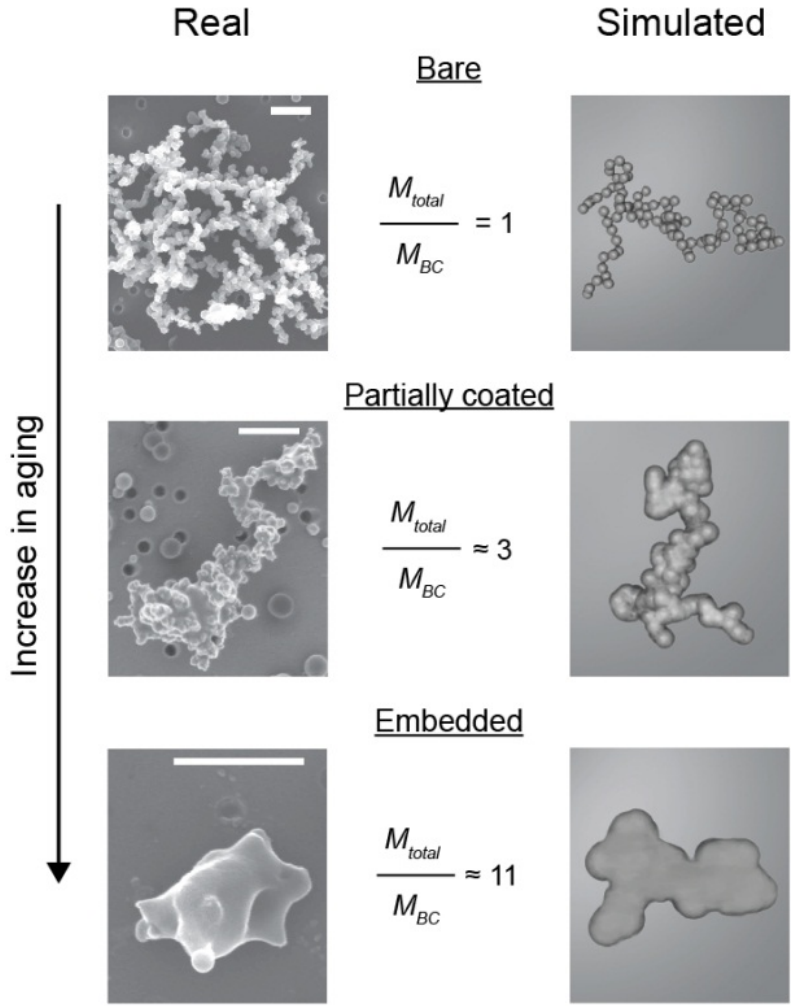

**Figure 1: Morphologies of internally mixed black carbon (BC) aggregates.** Three simulated aggregates from this study that were "cherry-picked" (shown in right column) demonstrates their close resemblance with real-world BC aerosol (left panel) corresponding to the three categories of internal mixing states as observed by China et al.[7]. The first row particles represent bare BC aggregate with point contacting monomers and an open fractal morphology; partially coated aggregates are shown in the second row; and thickly coated or embedded aggregates are displayed in the third row. The total particle mass to BC mass ratio is shown in the center for each class of particle.

In their bare state, BC aerosols appear as fluffy aggregates of carbon monomers–the number of monomers (N) scales with aggregate size (radius of gyration $R_g$) following a power-law relation:

$$N = k_0 \left(\frac{R_g}{a}\right)^{D_f} \qquad \text{[Eq.1]}$$

where $a$ is the monomer radius, $D_f$ is the mass fractal dimension and $k_0$ is a scaling prefactor[14]. Physically, an aggregate's $D_f$ describes its space-filling characteristic, while its shape anisotropy (stringiness) and monomer packing density are controlled by $k_0$[15]. The process of surface coating starts with filling in of the crevices between the monomers with non-refractory materials. If prolonged, this process could completely engulf the aggregate structure leaving only its thickened silhouette intact. To calculate the fractal parameters in Eq. 1 for a coated aggregate, we introduce an effective monomer radius $a'$, defined as $\frac{M_{total}}{M_{bare}} = \left(\frac{a'}{a}\right)^3$, which replaces $a$ in Eq. (1). This ensures the conservation of N between bare and coated aggregates. Aggregate $D_f$, determined using different techniques (see details in Methods), remains invariant at a fixed value of 1.8 with increasing coating mass, while $k_0$ scales as $k_0 = 1.34\left(\frac{M_{total}}{M_{bare}}\right)^{0.56}$. Details of this fractal formulation can be found in the recently published work by Heinson et al. [14].

We generated several hundred internally-mixed BC aggregates with $\frac{M_{total}}{M_{Bare}}$ ranging between 1 and 18 using our fractal aerosol model. For this study, we denoted $M_{bare} = M_{BC}$ and $M_{non\text{-}refractory\_coating} = M_{total} - M_{bare}$, and used the materials densities for BC and non-refractory coating to be 1.8 and 1.2 g/cm$^3$, respectively [16,17]. The $a$ for bare BC aggregates was fixed at 25 nm and the range of $N$ was varied between 10 and 250 following past field observations. We applied the dipole-dipole approximation electromagnetic theory [18] to compute the orientationally-averaged MAC$_{BC}$ and $E_{MAC_{BC}}$ for our simulated aggregates. Optical calculations were performed at three wavelengths $\lambda$=405nm, 532nm and 880nm, representing the near-UV, green, and red (near-IR) spectra of the incoming solar light. The complex indices of refraction for the BC monomers and refractory material coating were set at m=1.95-0.79i and 1.55-0i (no absorption), respectively [19-22]. This refractive index was chosen with the intention of making our findings applicable to broad-ranging scenarios of BC coated with OC and sulfates as found commonly occurring in the atmosphere.

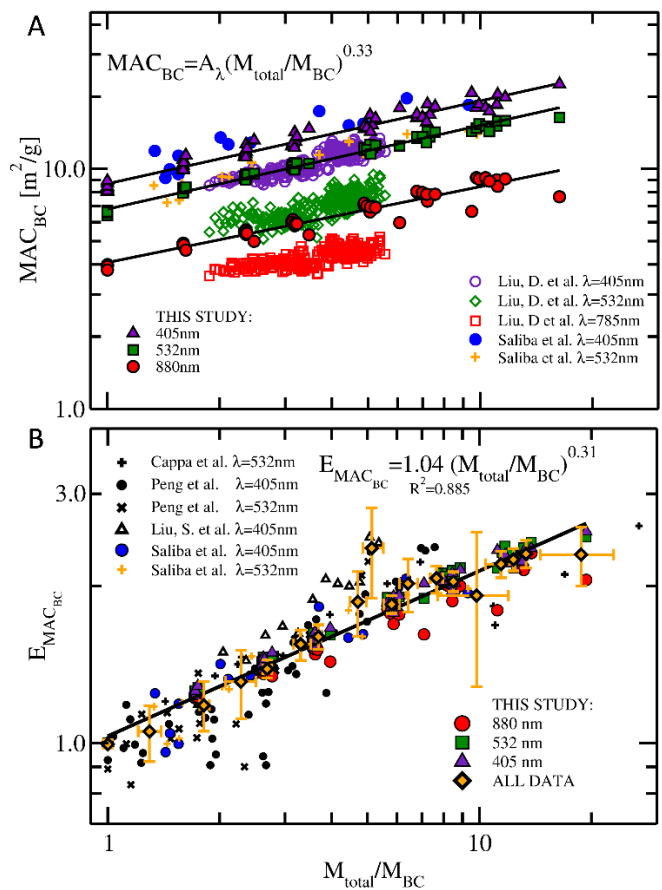

**Figure 2: The 1/3 Scaling Laws for BC Mass Absorption Cross section (MAC$_{BC}$) and Enhancement of MAC$_{BC}$ ($E_{MAC_{BC}}$).** The MAC vs. total coated aggregate mass divided by its bare BC mass ($M_{total}/M_{BC}$) at wavelengths $\lambda$=405nm, 532nm, 880nm is shown in (A). At all $\lambda$, the MAC scales with a power law exponent of 0.33±0.05. Our work compares well to observational findings on cook stove emissions done by Saliba et al. [23], while data from Liu, D. et al. [12] has smaller prefactor owing to integrated absorption measurements as opposed to single-particle, yet their trends are parallel to the other data sets. (B). The enhancement of the absorption $E_{MAC_{BC}}$ of internally mixed BC is plotted versus $\frac{M_{total}}{M_{BC}}$. The "ALL DATA" set of points represent the mean values of all dataset from this study; the error bars represent two standard deviations. Independent of $\lambda$, $E_{MAC_{BC}}$ for all data sets follows the same upward increasing trend with a power-law exponent of 0.31±0.05. The $E_{MAC_{BC}}$ estimates from our simulation agrees very well (regression coefficient $R^2$ = 0.885) with findings from past observational field studies carried out globally [8-10,12]. This agreement suggests a universal behavior for $E_{MAC_{BC}}$ for internally mixed BC at visible and near infrared wavelengths.

In figure 2A, we compare the calculated $MAC_{BC}$ versus $\frac{M_{total}}{M_{BC}}$ ratios for our aggregates with those reported by D. Liu et al. [12] and Saliba et al. [23]. D. Liu et al. performed a comprehensive set of laboratory and ambient experiments investigating optical properties of combustion aerosols generated from automotive diesel engines, and a large number of intensive open wood fires and fireworks across the UK. They did not measure $MAC_{BC}$ on a single particle level, instead estimated it from measurements of BC absorption coefficients–particle cross-sections integrated over a size distribution. Saliba and co-workers had coated freshly emitted BC aggregates from household cook stoves with secondary OC produced via the photo-oxidative ozonolysis of α-pinene. They measured $MAC_{BC}$ on a single particle level using a single-particle soot photometer. All datasets follow very well the scaling relationship $MAC_{BC} = A_\lambda \left(\frac{M_{total}}{M_{BC}}\right)^{0.33 \pm 0.05}$, with the prefactor $A_\lambda$ varying as $A_\lambda = 3.6\lambda^{-0.98}$. Our $MAC_{BC}$ for bare aggregates lie within the range of values reported for nascent soot by Bond et al. [1]. The variability in the prefactor $A_\lambda$ with $\frac{M_{total}}{M_{BC}}$ could be attributed to differences in refractive indices and mass densities of coating materials, distributions in monomer sizes, and errors involved in the different measurement techniques. This is a multidimensional parameter space that needs to be explored in detail as part of future studies. For instance, the variation in mass densities of OC coating materials alone could range from 0.64 to 1.65 g/cm$^3$ and 1.06 to 1.45 g/cm$^3$ for biogenic and anthropogenic emissions, respectively [24]. Similarly, the variation in real part of $m$ for OC aerosol could range from 1.36 to 1.66, while their imaginary part could have a non-zero and wavelength-dependent value [25].

In figure 2B, we show the universal scaling behavior of $E_{MAC_{BC}}$ as a function of $\frac{M_{total}}{M_{BC}}$ for all particles at the three wavelengths investigated in this study. Overlaid on our experimental results are observational datasets collected from different geographical regions. Peng et al.'s [10] data set, involving ambient carbonaceous aerosol collected in Houston (USA) and Beijing (China) followed by oxidation in an environmental chamber, is representative of primarily internally mixed BC occurring in ambient urban conditions of developed and developing countries. Liu, S. et al.'s [8] dataset is from fossil fuel and residential biofuel emissions in and nearby London (UK), and also includes inter-continentally transported, atmospherically processed particles. Cappa et al. [9] provided laboratory measurements of $E_{MAC_{BC}}$, which are in agreement with our trends; however, their unusual field findings on relatively smaller values of absorption enhancements even in the presence of substantial coatings present themselves as outliers. In addition to the datasets corresponding to different studies, we included in our analysis the mean values of all data, which in figure 2B is represented by the "ALL DATA" set of points. As evident from the figure, the datasets follow the scaling relation $E_{MAC_{BC}} = 1.04 \left( \frac{M_{total}}{M_{BC}} \right)^{0.31 \pm 0.05}$, independent of variations in wavelength, with high coefficients of regression. For partially coated aggregates, $E_{MAC_{BC}}$ ranged from 1.3 to 1.9 while the embedded aggregates had $E_{MAC_{BC}}$ ranging between 2.2 and 2.5.

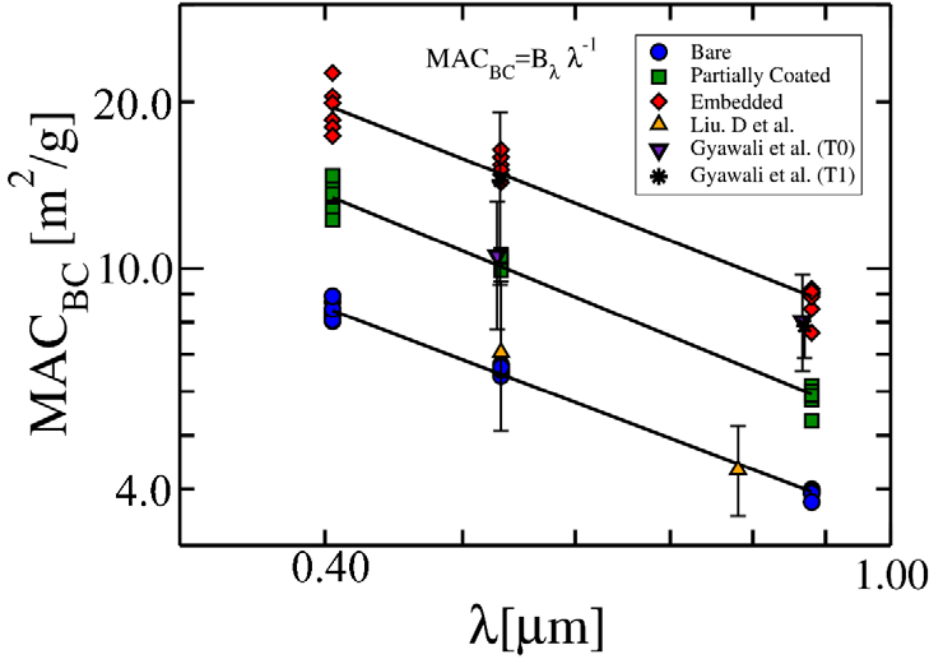

**Figure 3: Inverse wavelength scaling of BC Mass Absorption Cross section (MAC_BC).** The spectral response of MAC values for BC aggregates with varying degrees of mixing states are shown to follow an inverse functionality in wavelength (power law with exponent of -1). The fractal prefactor $B_\lambda$ ranges from 3.5 (bare) to 5.2 (partially coated) to 7.8 (embedded), and denotes the enhancement in MAC_BC through the phenomenological "lensing effect". Observational datasets collected over California, USA (Gyawali et al.[26]) and Manchester, UK (Liu, D et al.[12]) follow the scaling trends. Error bars indicate the standard deviation in the reported measurements.

The wavelength dependence of MAC_BC for our aggregates exhibits a constant $\lambda^{-1}$ power-law behavior (figure 3). Field datasets from Gyawali et al. [26] and Liu et al. [8] corroborates this constant scaling observation. Gyawali et al. characterized the evolution of multispectral optical properties of urban aerosols as they mixed with precursor gases and interacted with biogenic emissions during transportation to the forested Sierra Nevada foothills area in California. The power-law prefactor $B_\lambda$, however, increases as $3.6\left(\frac{M_{total}}{M_{BC}}\right)^{0.32}$ highlighting the enhanced focusing effect of the coating mass onto the BC core. The increased absorption cross-section of the BC core, owing to its fractal morphology, continues to exhibit Rayleigh optics behavior even at large values of $\frac{M_{total}}{M_{BC}}$ and $R_g$ (ca. 450 nm).

The core's fractal morphology dictates the particle's phase shift parameter $\rho$ to be always less than one, a necessary condition for the Rayleigh approximation to hold good [27,28]. $\rho$ is directly proportional to the volume fraction of monomers in an aggregate and the Lorentz-Lorenz factor involving the imaginary index of refraction (*16, 26*). It quantifies how much phase shift the incoming light waves encounter across an aggregate compared to that in the absence of the particle. For sub-micron size BC aggregates, $\rho$ scales with $Rg$ as $\rho \approx R_g^{-0.2}$ (See Supplementary Figs. S1 and S2) implying a decreasing $\rho$ with increasing aggregate size and its value remaining always less than 1. In the Rayleigh limit, the absorption cross-section of a BC aggregate could be simplistically calculated as $C_{abs} = N \cdot C_{abs,mono}$, where $C_{abs,mono} = 4\pi \frac{2\pi}{\lambda} a^3 Im\left(\frac{m^2-1}{m^2+2}\right)$ (where *Im* = imaginary part) is the absorption cross section of a monomer. The aggregate's MAC$_{BC}$, which is its $C_{abs}$ divided by aggregate mass, then becomes equal to the mass absorption cross-section of a monomer: $MAC_{BC} = \frac{N \cdot C_{abs,mono}}{N \cdot m_{mono}} = \frac{C_{abs,mono}}{m_{mono}} = MAC_{mono}$, thus explaining its inverse scaling functionality with wavelength.

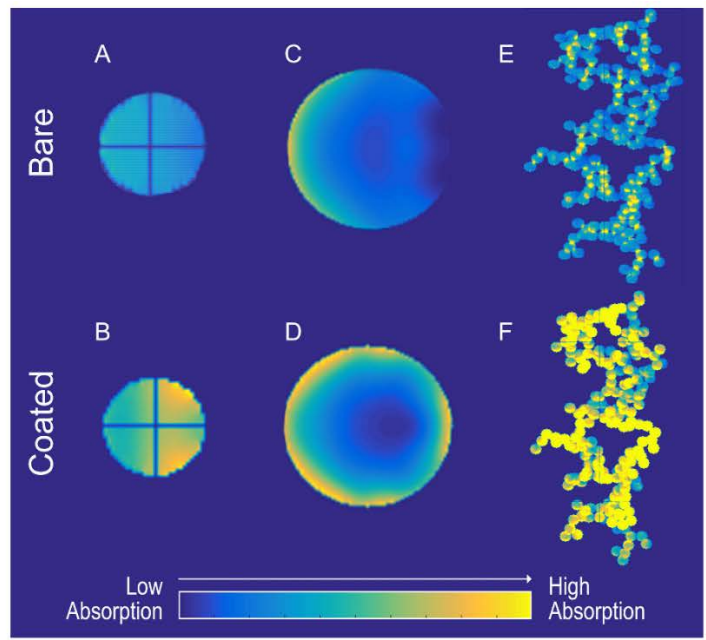

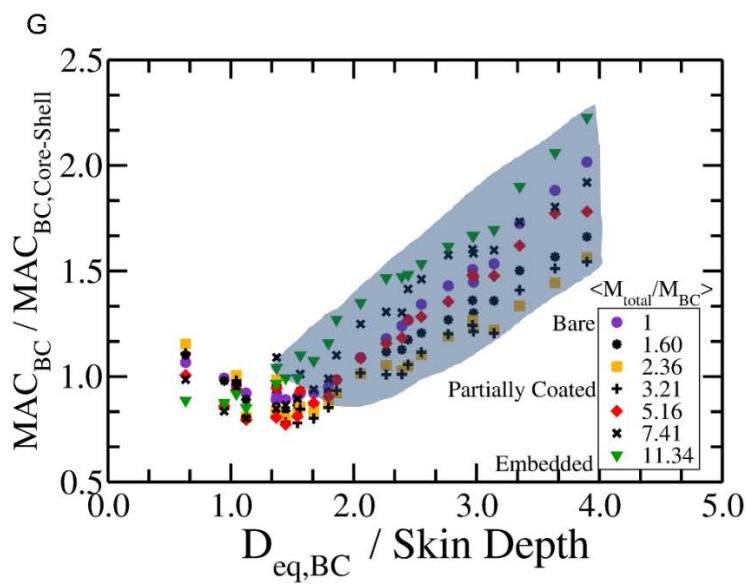

**Figure 4: Role of Particle Morphology in Light Absorption.** The strength of internal light absorption is shown for BC particles in their bare and heavily coated states (panels A-F). The coating material is non-absorbing and is therefore invisible in this representation. The incident light impacts the particles from the left. (A) BC sphere of diameter of D=80 nm; (B) Same sphere as (A) but having a 90 nm thick layer of coating; (C) BC sphere of D=300 nm; (D) Same sphere as (C) but with a 260 nm coating; and (E) and (F) represent BC aggregates having volume equivalence to spheres in (C) and (D), respectively. Particles (B), (D), and (F) are heavily coated which results in stronger absorption due to the "lensing effect." The diameter of spherical BC cores in (C) and (D) is larger than the absorption skin depth resulting in the interior of the particle being excluded from contributing to light absorption. The BC aggregate is porous in structure and therefore light penetrates completely into it allowing the entire volume to add to the total absorption (panels E-F). (G) The $MAC_{BC}$ ratios for coated aggregates to core-shell is plotted versus the mass equivalent sphere diameter $D_{eq,BC}$ divided by absorption skin depth. Significant disagreement is seen between the fractal and core-shell approximation models as $D_{eq,BC}$ becomes greater than the skin depth. As the equivalent spheres become larger than the absorption skin depth, the core-shell model underpredicts the $MAC_{BC}$ of the aggregates. The shaded region scales as $\frac{MAC_{BC}}{MAC_{BC,Core-Shell}} \propto \left(\frac{D_{eq,BC}}{Skin\ Depth}\right)^{0.75}$

We computed mass-equivalent core-shell structure spheres of our coated fractal aggregates and calculated their $MAC_{BC,core-shell}$ values using the Lorenz-Mie theory (figure 4). Climate models typically assume core-shell morphologies to account for internal mixing of BC and apply the computationally inexpensive Mie theory to calculate their optical properties [5,6,11]. Unlike fractal aggregate morphologies, $\rho$ for core-shell homogeneous spheres crosses over from the Rayleigh ($\rho < 1$) to the Geometric optics (GO) limit ($\rho > 1$) at diameters $100 - 300$ nm for $\lambda = 405 - 880$ nm, respectively. Once in the GO regime, the amount of light penetrating into a spherical BC core is determined by its optical skin depth [28,29]. This effect is clearly visible when mapping the internal absorption fields of an aggregate and its equivalent core-shell model (figure 4 A-F). As the core diameter $D_{eq,BC}$ increases, the ratio $\frac{D_{eq,BC}}{Skin\ Depth}$ increases linearly to values between 1 and 6 indicating significant screening of light by the core's interior.

In summary, we present the first empirical evidence of light absorption by atmospheric BC demonstrating universal patterns and simple scaling laws. Our major findings are summarized in Table 1. Scaling behaviors represent universal concepts that underlie non-equilibrium physical systems such as aerosols, and hold great promise to serve as computationally inexpensive parameterizations in climate models and satellite retrieval algorithms toward improving the accuracy of radiative forcing predictions. Use of the core-shell approximation has been suggested to introduce up to 50% uncertainty in modeled DRF by BC [8]. Via this study, we hope to convince the atmospheric community to refrain from using this approximation in the future. If use of this approximation is inevitable, then care must be taken to properly integrate the correction factor due to the optical skin depth of the BC core. Finally, we anticipate future laboratory and field studies to further refine these scaling laws, especially for unique case scenarios, such as BC core with absorbing coating materials, non-uniform distribution of coating mass on a core, and a core with $D_f$ approaching 3 [30]. As part of this study, we performed sensitivity analysis by assigning a weakly absorbing imaginary index ($\approx 5\text{x}10^{-2}$) to the OC core at 405 nm, representative of brown carbon. The deviations in scaling dependencies of $E_{MAC_{BC}}$ could be considered as negligible, but MAC$_{BC}$ values showed considerable deviations from the observed scaling dependencies.

| $Y$ | $Y_0$ | $S$ | $\beta$ |
|---|---|---|---|
| $E_{MAC_{BC}}$ | $1$ | $\dfrac{M_{total}}{M_{BC}}$ | $\sfrac{1}{3}$ |
| $MAC_{BC}$ | $3.6\lambda^{-1}$ | $\dfrac{M_{total}}{M_{BC}}$ | $\sfrac{1}{3}$ |
| | $3.6\left(\dfrac{M_{total}}{M_{BC}}\right)^{\sfrac{1}{3}}$ | $\lambda$ | $-1$ |
| $\dfrac{MAC_{BC}}{MAC_{BC,Core-Shell}}$ | $0.55 - 0.80$ | $\dfrac{D_{eq,BC}}{Skin\ Depth}$ | $\sfrac{3}{4}$ |

**Table 1:** Summary of Power-Law scaling relations of the form $Y = Y_0\ S^{\beta}$ for key light absorption parameters

**Acknowledgements:** This work was supported by the U.S. National Science Foundation (AGS-1455215, CBET-1511964, and AGS-PRF-1624814) and the National Aeronautics and Space Administration Radiation Sciences Program (NNX15AI66G) managed by Dr. Hal Maring. The authors thank Dantong Liu and James Allen of University of Manchester for sharing of the raw data from their field experiments.

**REFERENCES:**

[1]    T. Bond *et al.*, J. Geophys. Res., **118**, 5380 (2013).

[2]    V. Ramanathan and G. Carmichael, Nat. Geosci. **1**, 221 (2008).

[3]    Ö. Gustafsson and V. Ramanathan, Proceedings of the National Academy of Sciences **113**, 4243 (2016).

[4]    O. Boucher *et al.*, Proceedings of the National Academy of Sciences, 201607005 (2016).

[5]    M. G. Flanner, C. S. Zender, J. T. Randerson, and P. J. Rasch, J. Geophys. Res., **112** (2007).

[6]    Q. Wang *et al.*, J. Geophys. Res., **119**, 195 (2014).

[7]    S. China, C. Mazzoleni, K. Gorkowski, A. C. Aiken, and M. K. Dubey, Nature Communications **4** (2013).

[8]    S. Liu *et al.*, Nature communications **6** (2015).

[9]    C. D. Cappa *et al.*, Science **337**, 1078 (2012).

[10]   J. Peng *et al.*, Proceedings of the National Academy of Sciences **113**, 4266 (2016).

[11]   M. Z. Jacobson, Nature **409**, 695 (2001).

[12]   D. Liu *et al.*, Nat. Geosci. **10**, 184 (2017).

[13]   S. H. Chung and J. H. Seinfeld, J. Geophys. Res., **110** (2005).

[14]   W. R. Heinson, P. Liu, and R. K. Chakrabarty, Aerosol Sci. Technol. **51**, 12 (2017).

[15]   W. Heinson, C. Sorensen, and A. Chakrabarti, J. Colloid Interface Sci. **375**, 65 (2012).

[16]   B. J. Turpin and H.-J. Lim, Aerosol Sci. Technol. **35**, 602 (2001).

[17]   M. Hess, P. Koepke, and I. Schult, Bulletin of the American Meteorological Society **79**, 831 (1998).

[18]   M. A. Yurkin and A. G. Hoekstra, Journal of Quantitative Spectroscopy and Radiative Transfer **112**, 2234 (2011).

[19]   T. Bond and R. Bergstrom, Aerosol Sci. Technol. **40**, 27 (2006).

[20]   C. H. Jung, H. J. Shin, J. Y. Lee, and Y. P. Kim, Atmosphere **7**, 65 (2016).

[21]   H. Horvath, Atmos. Environ. **27**, 293 (1993).

[22]   P. DeCarlo *et al.*, Atmos. Chem. Phys. **7**, 18269 (2007).

[23]   G. Saliba *et al.*, Aerosol Sci. Technol. **50**, 1264 (2016).

[24]   M. Hallquist *et al.*, Atmos. Chem. Phys. **9**, 5155 (2009).

[25]   T. Moise, J. M. Flores, and Y. Rudich, Chem. Rev. (Washington, DC, U. S.) **115**, 4400 (2015).

[26]   M. Gyawali *et al.*, Atmosphere **8**, 217 (2017).

[27]   C. M. Sorensen, Aerosol Sci. Technol. **35**, 648 (2001).

[28]   H. van de Hulst, New York: Dover, 1981  (1981).

[29]   G. Wang, A. Chakrabarti, and C. M. Sorensen, JOSA A **32**, 1231 (2015).

[30]   Y. Wang *et al.*, Environmental Science & Technology Letters **4**, 487 (2017).

**Title: Scaling laws for Light Absorption by Atmospheric Black Carbon Aerosol**

**Authors:** Rajan K. Chakrabarty[1,2*] and William R. Heinson[1]

**Affiliations:**

[1]Center for Aerosol Science and Engineering, Department of Energy, Environmental and Chemical Engineering, Washington University in St. Louis, Missouri – 63130, USA.

[2]McDonnell Center for the Space Sciences, Washington University in St. Louis, Missouri – 63130, USA.

* Correspondence to*: chakrabarty@wustl.edu*

# Supplementary Information

**Diffusion limited cluster-cluster aggregation (DLCA).** Bare aggregates were generated using an off-lattice, DLCA algorithm [1,2]. Initially $10^7$ spherical monomers of radius $a = 25$ nm were randomly placed into a three-dimensional simulation box. As DLCA starts, the number of aggregates $N_c$ including lone monomers is counted. An aggregate is randomly chosen and simulation time is incremented by $N_c^{-1}$. The probability that an aggregate moves is inversely proportional to the aggregate's radius of gyration and is normalized to insure the monomers will always move upon selection. If the aggregate moves, it travels randomly one monomer diameter $2a$. When two aggregates collide, they irreversibly stick and $N_c$ is decremented by 1. Results are applicable in the continuum limit where the frictional drag is given by the Stokes–Einstein expression with a drag proportional to the radius of gyration.

**Coating of DLCA aggregates.** After a population of aggregates was generated, the process of coating was initiated. The coating algorithm first discretized the simulation space into high-resolution cubic lattices of equal sub-volumes. A bare DLCA aggregate was placed in the simulation space, which resulted in the sub-volumes being either filled with portions of the aggregate or left empty. Empty sub-volumes that were on or bordering the aggregate surface were identified and filled with the coating material. Next, the algorithm checked if the desired coating thickness was reached. If not, the process of identifying and filling empty sub-volumes on the aggregate surface was repeated.

**Determination of fractal dimension $D_f$ and prefactor $k_0$.** At each coating thickness, the average $a'$ was found and used in place of bare monomer $a$ in Eq. (1). Number of monomers $N$ in Eq. (1) was calculated by dividing the aggregate total mass by the mass of the effective monomer with radius $a'$. The radius of gyration is equivalent to the root mean square distance of the sub-volumes that make up the coated aggregate:

$$R_g^2 = \frac{1}{N_{SV}} \sum_{i=1}^{N_{SV}} (r_i - r_{mean})^2 \qquad \text{[Eq. S1]}$$

where $r_{mean}$ is the mean position of the sub-volumes, $r_i$ is the position of the $i^{th}$ sub-volume, and $N_{SV}$ is the total number of sub-volumes in an aggregate. The replacement of $a$ with $a'$ conserves monomer count $N$ for a given aggregate as coating material is added. Plotting N versus $R_g$ for the ensemble of aggregates yielded $D_f$ and $k_0$.

To find the $D_f$ of single aggregate the reciprocal space structure factor method was employed. The reciprocal space structure factor of the aggregate is described the Fourier transform squared of the sub-volume coordinates:

$$S(q) = \frac{1}{N_{SV}} \left| \sum_{i=1}^{N_{SV}} e^{-iq \cdot r_i} \right|^2 \qquad [\text{Eq. S2}]$$

where q is the Fourier variable with units of inverse length. After manipulation of Eq. S2, details of which can be found in[1,3], the structure factor can be written as:

$$S(q) = \frac{k_0}{a'^{D_f}} q^{-D_f} \qquad [\text{Eq. S3}]$$

From the S(q) plot, $D_f$ was calculated.

**The discrete dipole approximation (DDA).** We calculated the optical properties of our simulated aggregates and their internally mixed counter parts by using the discrete dipole approximation DDA [4]. The advantages of this method are that it is a numerically exact solution based on the Maxwell equations and it is versatile in its treatment of arbitrary particle shapes in different mixed states. The DDA implementation used in this work was the parallelized ADDA 1.3b4 code, publicly available at https://github.com/adda-team/adda. All runs had a dipole resolution of 12 dipoles per bare monomer diameter and were well within often cited of $k|m|d<1$ where $d$ is the dipole length [5]. Furthermore, the smallest optical skin depth for any simulated particle was $20d$ again well within the validity criteria in[6]. To further ensure the accuracy of our DDA runs, we compared core-shell model results from the Mie solutions and the DDA method. Comparisons of a lone monomer and the volume equivalent sphere of the largest simulated aggregate were done for both the bare and heavily coated cases. At both size extremes, R = 25nm for the lone monomer and $R_{eq}$ = 420nm for the heavily coated equivalent radius sphere, the error between the numerical methods was always less than 10 percent. It should also be noted that previous work has shown that spherical particles produce the largest errors in DDA and the error for aggregates is expected to be much smaller [5].

**Core-Shell model and calculation of skin-depth and effective refractive index.** For each aggregate, either in bare or internally mixed states, a volume equivalent sphere with a BC core and OC shell was created. Next, their optical properties were calculated using both DDA and Lorenz-Mie theories [4,7]. Absorbing material such as BC attenuates incident light exponentially according to the Beer-Lambert Law where the intensity of the incident light penetrating a particle is described by:

$$I(z) = I_0 e^{-\alpha z} \qquad [\text{Eq. S4}]$$

where $I_0$ is the intensity of the light at the particle surface and $\alpha^{-1}$ is the characteristic penetration length. The skin depth of an absorbing material is defined as the depth at which the intensity of the incident light falls to $1/e^2$ of its surface value and is related to the complex refractive index $m$ and incident wavelength $\lambda$ by:

$$skin\ depth = \frac{\lambda}{2\pi \cdot Im(m)} \qquad [Eq.\ S5]$$

where $Im(m)$ is imagery part of the complex index of refraction. The equivalent core-shell spheres quickly grow larger than the skin depth which results in a drop in absorption cross-section as compared to their aggregate counterparts.

On the other hand, the porous nature of aggregates necessitates the need to apply effective medium theory. A porous object would have a significant different $m_{eff}$, compared to an equivalent sphere with homogeneous index $m$. The Maxwell–Garnet effective medium theory[8] conveniently provides a way to calculate the $m_{eff}$ of inhomogeneous particles. In the case of void-filled aggregates, the theory calculates $m_{eff}$ using the relation:

$$f_v \left(\frac{m^2-1}{m^2+2}\right) = \left(\frac{m_{eff}^2-1}{m_{eff}^2+2}\right) \qquad [Eq.\ S6]$$

where $f_v$ is the volume fraction of the BC aggregates. Since DLCA aggregates scale with fractal dimension $D_f = 1.8$ their $m_{eff}$ continues to decrease and in turn skin depth increases as the aggregates grow larger. The $\rho$ parameter which describes an object's transition from Rayleigh to geometric optics is found by simply multiplying the above equation by the aggregate size parameter.

**Volume fraction of aggregates.** The volume fraction $f_v$ of aggregates is found by first aligning the aggregates along their principle axes and constructing an encapsulating ellipsoid. The principle axes were found by calculating the eigenvectors of the inertia tensor $T$ [3,6]. For an aggregate of $N$ discrete monomers the inertia tensor is given by:

$$T = \sum_{i=1}^{N} \begin{pmatrix} y_i^2 + z_i^2 & -x_i y_i & -x_i z_i \\ -x_i y_i & x_i^2 + z_i^2 & -y_i z_i \\ -x_i z_i & -y_i z_i & x_i^2 + y_i^2 \end{pmatrix}$$

The volume fraction is then defined as the ratio total monomer volume to the encapsulating ellipsoid volume:

$$f_v = \frac{8Na^3}{L_1 L_2 L_3}$$

where $L_i$ are the lengths of aggregate along the principle axes.

**Internal Absorption Strength.** DDA algorithms map a particle into an array of small subvolumes that act as point dipoles. All optical properties reported by DDA come from integration over the dipoles' final calculated values. The absorption cross-section, $C_{abs}$ is calculated from the dipoles using the following relationship:

$$C_{abs} = \frac{4\pi k}{Re(m)} \sum_i Im(P_i \cdot E_i^*)$$

where $k = 2\pi/\lambda$, $P_i$ and $E_i$ are the $i^{th}$ dipole polarization and electric field respectively. $Im(P_i \cdot E_i^*)$ is imaginary part of the dot product between the polarization and the complex conjugate of the electric field and can be thought of the absorption strength of the $i^{th}$ dipole. To simplify $Im(P_i \cdot E_i^*)$ we start with the expression for that relates the dipole polarization is to the electric:

$$P_i = \varepsilon_0 \chi E_i$$

where the susceptibility $\chi$ is related to the index of refraction $m$ by

$$m^2 = 1 + \chi$$

Given the electric field of the $i^{th}$ dipole as

$$E_i = E_{0,i} \, exp(i \, \boldsymbol{k} \cdot \boldsymbol{r})$$

The dot product $P_i \cdot E_i^*$ becomes

$$P_i \cdot E_i^* = \varepsilon_0(m^2 - 1)E_{0,i} \, exp(i \, \boldsymbol{k} \cdot \boldsymbol{r})E_{0,i} \, exp(-i \, \boldsymbol{k} \cdot \boldsymbol{r})$$

which simplifies to

$$P_i \cdot E_i^* = \varepsilon_0[(n^2 + \kappa^2 - 1) + i2n\kappa]E_{0,i}^2$$

We arrive at the final expression for a dipole's absorption strength:

$$Im(P_i \cdot E_i^*) = 2\varepsilon_0 n\kappa E_{0,i}^2$$

**Figures**

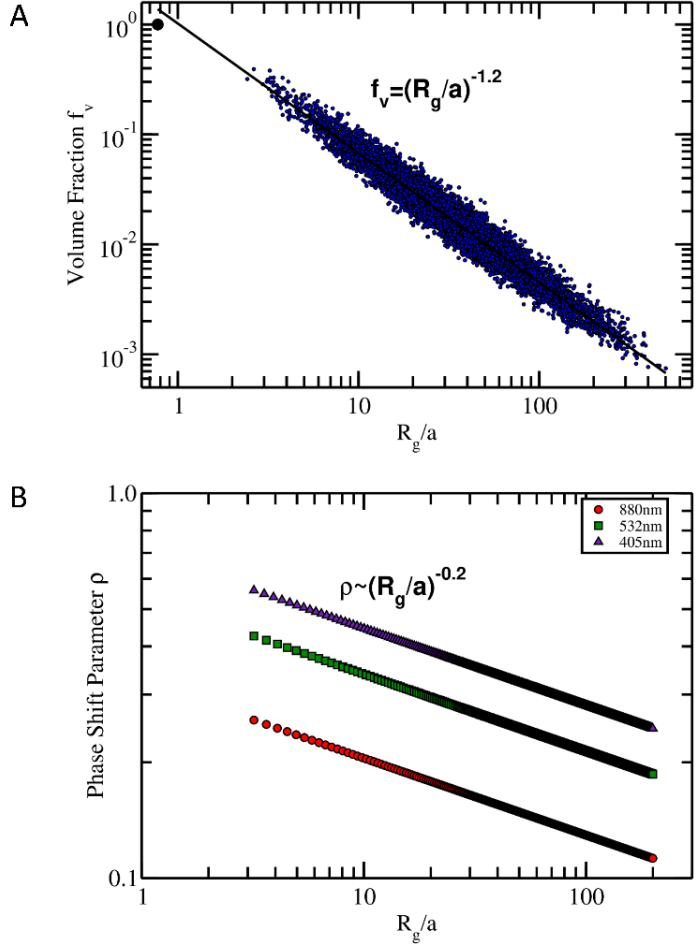

Fig. S1: **(A)** The volume fraction, $f_v$ of BC is calculated by encapsulating the aggregates in the smallest enclosing ellipsoid and then taking the ratio of occupied volume to total volume. The volume fraction scales as $d - D_f$, where $d$ is the spatial dimension and $D_f$ is the aggregate fractal dimension **(B)** By application of the Maxwell-Garnett effective medium theory, the measure of how coupled the internal electric field of an aggerate is found to be $\rho = 2Xf_v \left| \frac{m^2-1}{m^2+2} \right|$ where $X$ is the aggregate size parameter. If $\rho < 1$ the particle will be in the Rayleigh–Debye–Gans (RDG) regime and all optical properties will scale accordingly. For aggregates with small fractal dimension $D_f \leq 2$, $\rho$ will only decrease with size as is the case with DLCA aggregates of $D_f = 1.8$.

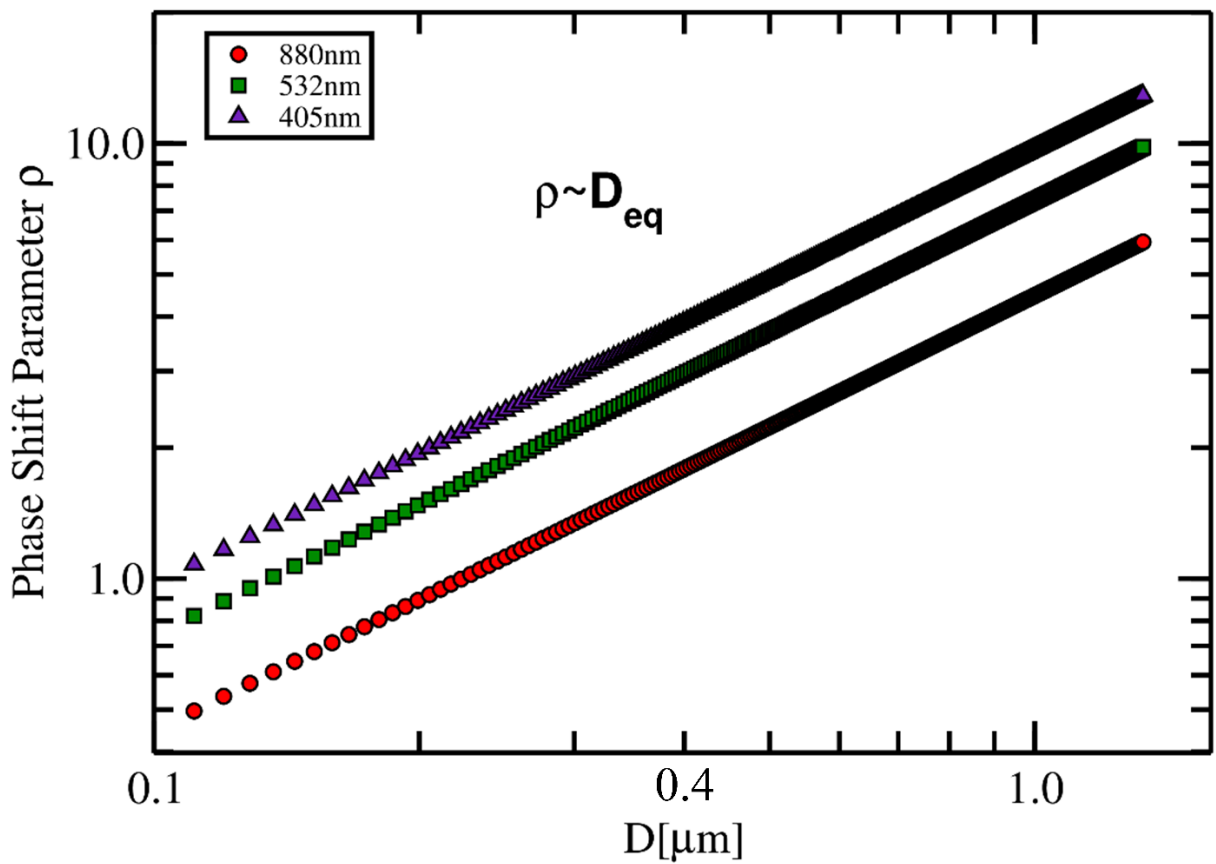

**Fig. S2:** Contrary to figure S1B, spheres of equivalent mass will increase in $\rho$ as they grow in diameter due to spheres having $D_f = d = 3$. Consequently, the disagreement between coated aggregate and core-shell models increase with size.

**References**

[1]    W. R. Heinson, P. Liu, and R. K. Chakrabarty, Aerosol Sci. Technol. 51, 12 (2017).

[2]    P. Meakin, J. Colloid Interface Sci. 102, 505 (1984).

[3]    W. Heinson, C. Sorensen, and A. Chakrabarti, J. Colloid Interface Sci. 375, 65 (2012).

[4]    M. A. Yurkin and A. G. Hoekstra, Journal of Quantitative Spectroscopy and Radiative Transfer 112, 2234 (2011).

[5]    E. Zubko *et al.*, Appl. Opt. 49, 1267 (2010).

[6]    D. Fry, A. Mohammad, A. Chakrabarti, and C. Sorensen, Langmuir 20, 7871 (2004).

[7]    C. F. Bohren and D. R. Huffman, J Wiley & Sons, New York  (1983).

[8]    N. V. Voshchinnikov, G. Videen, and T. Henning, Appl. Opt. 46, 4065 (2007).